\documentclass[conference]{IEEEtran}
\usepackage[ruled,vlined]{algorithm2e}
\usepackage{subcaption}
\usepackage{booktabs}
\usepackage{tabularx}
\usepackage{arydshln}
\usepackage{cite}
\usepackage{amsmath,amssymb,amsfonts,amsthm}
\newtheorem{definition}{Definition}
\newtheorem{remark}{Remark}
\newtheorem{lemma}{Lemma}
\newtheorem{theorem}{Theorem}

\usepackage{algorithmic}
\usepackage{graphicx}
\usepackage{textcomp}
\usepackage{xcolor}
\def\BibTeX{{\rm B\kern-.05em{\sc i\kern-.025em b}\kern-.08em
    T\kern-.1667em\lower.7ex\hbox{E}\kern-.125emX}}
\usepackage{hyperref}

\hypersetup{colorlinks,breaklinks,
  urlcolor=black,
  linkcolor=black,
  citecolor=black,
}
\usepackage{float}
\usepackage{listings}
\usepackage{newfloat,caption}
\DeclareFloatingEnvironment[fileext=frm,placement={!ht},name=Listing,within=section]{listing}
\IEEEoverridecommandlockouts
\begin{document}

\title{Identifying Malicious Players in GWAP-based Disaster Monitoring Crowdsourcing System*\\
{}
\thanks{
  \footnotesize \textsuperscript{*}
  This research has been supported by the Bavarian IUK Program (IUK-1805-0004//IUK577/002) 
  and Sichuan Science and Technology Program (2019YFH0055).
}
}

\author{
\IEEEauthorblockN{Changkun Ou}
\IEEEauthorblockA{
\small
Institute of Computer Science \\
University of Munich\\
Munich, Germany \\
changkun.ou@ifi.lmu.de
}
\and
\IEEEauthorblockN{Yifei Zhan}
\IEEEauthorblockA{
\small
Institute of Computer Science \\
University of Munich\\
Munich, Germany \\
yifei.zhan@campus.lmu.de
}
\and
\IEEEauthorblockN{Yaxi Chen \textsuperscript{1,2}}
\IEEEauthorblockA{
\small
1. The Key Laboratory for Computer Systems of \\
State Ethnic Affairs Commission \\
2. School of Computer Science and Technology\\
Southwest Minzu University\\
Chengdu, China \\
yaxichen@swun.cn
}
}

\maketitle

\begin{abstract}
Disaster monitoring is challenging due to 
the lake of infrastructures in monitoring areas.
Based on the theory of Game-With-A-Purpose (GWAP), this paper contributes to a novel large-scale crowdsourcing disaster monitoring system.
The system analyzes tagged satellite pictures from anonymous players,
and then reports aggregated and evaluated monitoring results to its stakeholders.
An algorithm based on directed graph centralities is presented to address
the core issues of malicious user detection
and disaster level calculation. Our method can be easily applied in other human computation systems.
In the end, some issues with possible solutions are discussed for our future work.
\end{abstract}


\section{Introduction}

\IEEEPARstart{M}ANY Non-Profit Organizations (NPOs) such as the United Nations Children's Fund (UNICEF)
provide \cite{unicef1994state} humanitarian assistance in developing contries.
The largest challenges for these organizations are those
unreachable zones \cite{unicef2017report} where the real time war situation or disaster level are extremely difficult to be derived.
Lack of sufficient local infrastructures, disasters can only be monitored from the sky level.
Satellite sensors are widely deployed in order to report
images of monitoring areas \cite{zhang2002flood}.

Nowadays automatic disaster monitoring has not achieve satisfying success while highly costly manual methods
cannot satisfy real-time requirements.
Therefore, the fields of human computation and crowdsourcing are investigating methods
to harvest crowd wisdom.
GWAP is one representative theory which convert time- and energy-consuming image processing
problems into games in which players are motivated to contribute.
Inspired by this theory, we present 
a novel large-scale crowdsourcing disaster monitoring system. 
The system analyzes tagged satellite pictures from players 
and then calculate the disaster level automatically.
An algorithm based on directed graph centralities is presented to address the core issues of malicious player detection
as well as disaster level calculation. 
Out method can be applied to other human computation systems in general.
As justification, the mathematical correctness of the system is proved.
In the end, we also discusse some limitations and relevant solutions for the future work.

\section{Related works}

\begin{figure*}[!ht]
  \centering
  \includegraphics[width=\textwidth]{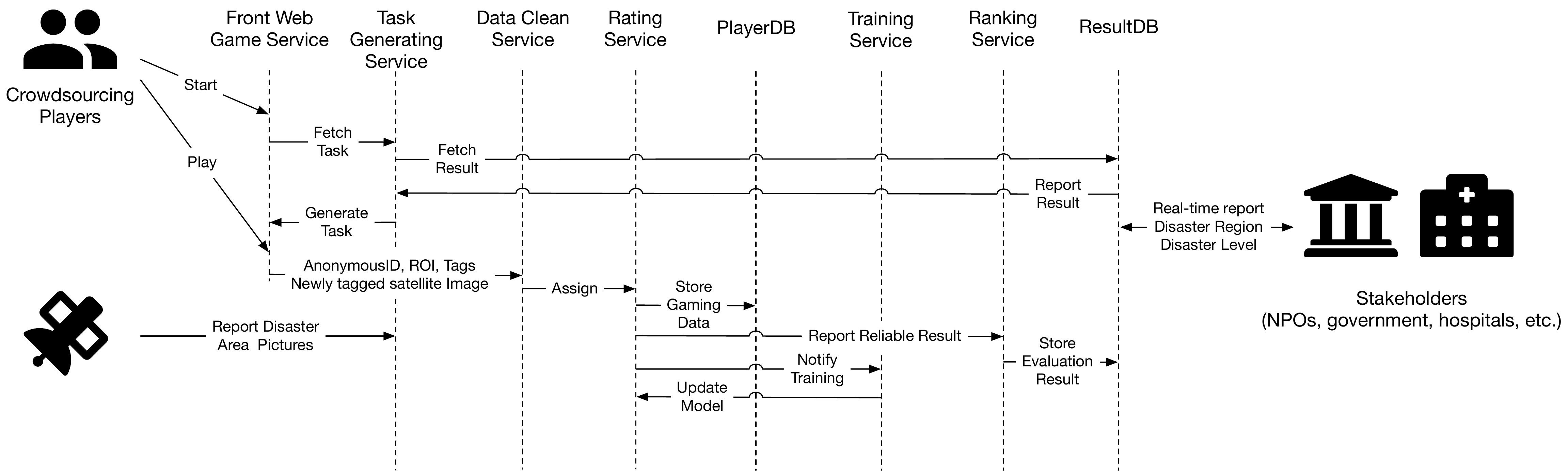}
  \caption{System architecture: the system is designed to cooperate with multiple microservices,
  the most critical components are task generating service, the rating service and ranking service. 
  The task generating service mixes reliable resulted images and new reported satellite images
  to generate player's tasks. 
  The rating service and ranking service are based on the PRM and DEM model respectively.
  }
  \label{fig:arch}
\end{figure*}

Human computation system is a paradigm for utilizing 
the human processing power to solve problems that 
neither computers or humans can solve independently \cite{quinn2009taxonomy, luisvon-hc}.
Most of the human computation systems
can be seen \cite{quinn2011human} as crowdsourcing trade,
which rely on the \emph{wisdom of crowds}.
Surowiecki claimed \cite{mennis2006wisdom} four critical properties of wisdom of crowds:
diversity of opinion, independence, decentralization and aggregation.
Oinas-Kukkonen further concluded \cite{oinas2008network} the theoretical foundation of wisdom of crowds based on
network analysis. 
For instance, PageRank was first proposed by 
Lary Page \cite{page1999pagerank} and 
applied to social network analysis \cite{bonacich2001eigenvector}.
It is commonly used for expressing the stability of 
physical systems and the relative importance, 
so-called centralities, of the nodes of a network.
PageRank fulfil the four condition of a wisdom of crowd mentioned above.

The fundamental theory for this paper is Game-With-A-Purpose (GWAP),
which involves game theory \cite{Jain:2008:GAG:1504941.1504990} in
human computation systems \cite{von2004labeling, von2006games}.
It outsourced within a computational process to 
humans in an entertaining way, namely gamification, and recently considered as
the power of addressing large-scale data labeling costs in
machine learning research\cite{daniel2018quality, ertel2018introduction, ardalan2018large}.
Nevertheless, the data collection mechanisms for a game is 
variety that should be considered in a proper way \cite{von2008designing}.
In long-term research, ESP \cite{von2004labeling}, and ARTigo \cite{wieser2013artigo} 
have verified through years of operations that human inputs are valuable and meaningful,
and the most important two challenges in GWAP systems are game incentivization 
and malicious player detection.

Unfortunately, these existing representative GWAP-based human computation systems
have the following issues:
(1) They require two online players competing with each other, 
which may harm the degree of playability and even meet troubles when lacking of players.
(2) They only use the most commonly appeared tags that cannot prevent
massive malicious players attacking the system and providing meaningless tags.
However, manually managing the tag database is not feasible due to the high cost of 
human labor and the inevitable issue of system cold start.
In order to deal with the lack of players, our system turns multiplayers-required game into 
game between new players and existing reliable players. 
Furthermore, a malicious player detection algorithm based on directed graph centralities 
is proposed which requires only one single reliable players to avoid the issue of cold start.

\section{Design and Models}
\label{sec:design}

In this section, we describe the overall design and proposed models in detail.
First, we propose the system architecture and
specify the most critical components:
\emph{player task generator (PTG)}, 
\emph{player rating model (PRM)} 
as well as \emph{disaster evaluation model (DEM)}.
With these components, the disaster monitoring system can handle
the common issues in human computation system, 
such as system cold start and malicious player detection. 
It is also expandable, portable and can be easily applied to any other similar 
human computation systems.

\subsection{System Architecture and Functionalities}

Figure \ref{fig:arch} illustrates the architecture of our disaster monitoring system.
The system databases are composed of two different type of databases. 
The player database (PlayerDB) stores gaming data including
the player's property and raw tagging inputs.
The other database is called ResultDB where persistents the reliable players' inputs
that rated by our rating service.
The overall data flow can be described as following:

\paragraph*{Step 1} Player task generation: 
  The \emph{PTG}
  mixes the reliable gaming results from ResultDB
  and new reported images from satellite,
  and then assigns them to the future players.
\paragraph*{Step 2} Malicious player detection: 
  A reliable player requires to pass the malicious detection algorithm 
  (see Algorithm \ref{algo:malicious})
  embedded in the \emph{PRM}. 
  Then the system will mark 
  all the results from this player as reliable
  and then send them to the ranking service.
\paragraph*{Step 3} Disaster level evaluation: 
  the system reuses the reliable players' inputs 
  into \emph{DEM} that embedded
  in ranking service and
  calculates the disaster level of the monitoring region
  then persistents it in ResultDB.

After these three major steps, a disaster level report can be retrieved from ResultDB.

\begin{figure}
  \centering
  \begin{subfigure}[h]{0.32\linewidth}
    \centering
    \includegraphics[height=18mm]{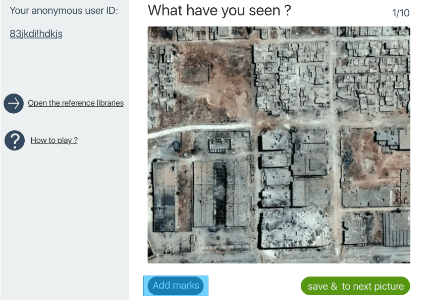}
    \caption{}\label{fig:player0}
  \end{subfigure}
  \begin{subfigure}[h]{0.32\linewidth}
    \centering
    \includegraphics[height=18mm]{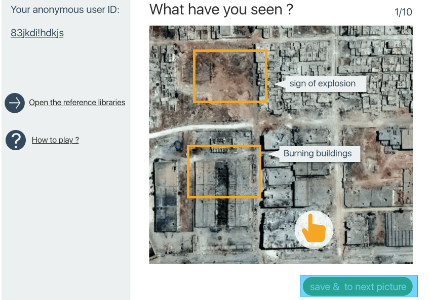}
    \caption{}\label{fig:player2}
  \end{subfigure}
  \begin{subfigure}[h]{0.32\linewidth}
    \centering
    \includegraphics[height=18mm]{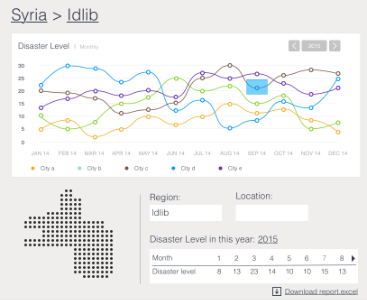}
    \caption{}\label{fig:stakeholder3}
  \end{subfigure}
  \label{fig:systempanel}
  \caption{
    System interface.
    a) Player game panel overview;
    b) Multi-tags selection for selected areas;
    c) Disaster level report in stakeholder view.
    Satellite images are taken from \cite{satellite-photo-1}.
  }
\end{figure}

In our game,
a player can execute infinity rounds of tasks, and
each single round of task contains $n$ image tagging tasks.
In one task, the player is asked to tag $n$ images (see Figure \ref{fig:player0}).
The player needs to draw a rectangle to select an area
where a sign of danger or damage (such as fire or explosion) is discovered.
System-suggested tags will then pop up and the player can select 
relevant ones by simply clicking on them
(see Figure \ref{fig:player2}). The player can also input new tags.
The system analyzes the user input and creates a disaster level report (see Figure \ref{fig:stakeholder3}) 
for this region which can be used by NGOs and governments.

\subsection{Preliminaries}

To describe and establish our models, 
we describe a few basic definitions in this subsection.

\begin{definition}
\label{def:roi}
The region of interests (ROI) is an indicator that represents
player-selected two-dimentional region.
The $i$-th ROI from player $p$ in image $k$ at image creation time $t$ is denoted by $ROI_{p,i,k,t}$.
\end{definition}

\begin{figure}[htp]
  \centering
  \includegraphics[width=\linewidth]{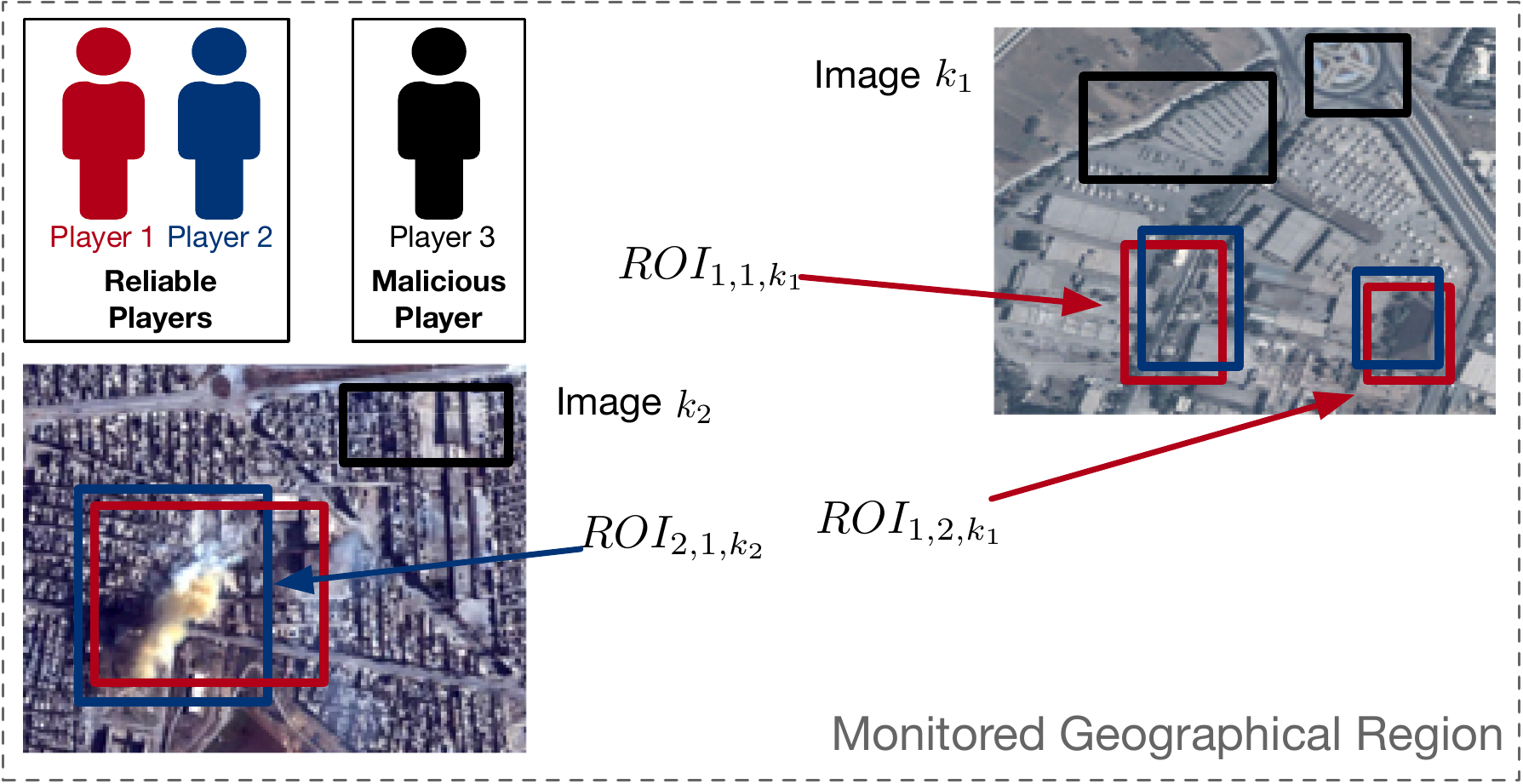}
  \caption{Examples of ROIs in the disaster monitoring system:
  In this figure, reliable players (red and blue) draw rectangles to indicate area with disaster, however a player that does not cooperate with the game (black)
  selects different ROIs. The $ROI_{p,i,k}$ indicates the $i$-th ROI from player $p$ in image $k$.}
  \label{fig:roi}
\end{figure}

Considering image $k$ implies its creation time $t$ (an image always contains its creation time), 
for convenience, \emph{$ROI_{p,i,k,t}$ is simplified as $ROI_{p,i,k}$}.
For instance, Figure \ref{fig:roi} shows some examples of ROIs in different images.

\begin{remark}\emph{
The definition is designed for our database scheme, which includes PlayerDB and ResultDB, 
and their data schemes are illustrated in Listing \ref{lst:playerdb}
and Listing \ref{lst:resultdb}.
See Appendix \ref{appendix:scheme} for more descriptions.
}\end{remark}

As we discussed in Section \ref{sec:design},
\emph{each tag can only be selected once, and players are allowed to input 
new tags for the selected ROIs.} 
Then, We define the ROI tag vector for the model:

\begin{definition}
\label{def:tagv}
Assuming the database stores $n$ different tags $g_1$, $g_2$, ..., $g_n$ for a certain image $k$,
the tag vector $\mathbf{T}_{p, i, k}$ of $ROI_{p, i, k}$ (the $i$-th ROI in image $k$ of player $p$) is a vector that 
is denoted by the following formula:
\begin{equation}
  \mathbf{T}_{p, i, k} = (|g_1|, |g_2|, ..., |g_n|)^\top
\end{equation}
where $g_l$ is the $l$-th tag where $l=1, 2, ..., n$, 
$|g_l|$ is the count of $g_l$ in a player task object, and $n$ equals to the number of tags.
\end{definition}

\begin{remark}\emph{
Since each tag can only be selected once, the components of tag vector is either 1 or 0.
This definition performs a popular data preprocessing technique, 
which called One-Hot Encoding trick \cite{wu2012foundations, liu2002discretization}.
}\end{remark}

\begin{remark}\emph{
For instance, for a certain image $k$,
5 different tags $g_1$, $g_2$, $g_3$, $g_4$, $g_5$ were input
by our game player.
Assuming player $p$ selects the first ROI and inputs tags for $ROI_{p, 1, k}$: 
$\{g_1$, $g_2$, $g_3$, $g_4\}$, 
and player $q$ selects the first ROI and inputs tags for $ROI_{q, 1, k}$:
$\{g_1, g_3, g_4, g_5\}$. 
Then tag vector $\mathbf{T}_{p, 1, k}$ of $ROI_{p, 1, k}$ is $(1, 1, 1, 1, 0)^\top$ and tag vector
$\mathbf{T}_{q, 1, k}$ of $ROI_{q, 1, k}$ is $(1, 0, 1, 1, 1)^\top$.
}\end{remark}

\subsection{Player Task Generator}

The \emph{PTG} creates task images by combining images from satellite and ResultDB.
A player task contains $2n$ different images in random order, in which $n$ images are
untagged new satellite images and other $n$ images are tagged images from ResultDB,
PTG thus contains two generating steps:

\paragraph*{Step 1} PTG splits a monitoring region into small pieces of images, 
assigning a unique identifier for each piece 
(The reason is discussed in Section \ref{sec:ethical}).

\paragraph*{Step 2} PTG retrieves tagged images from ResultDB, then combines
these two types of images to create a task for a new upcoming player.

\subsection{Player Rating Model}

The PRM is responsible for detecting malicious players.
We convey the basic idea of centralities of a network
and use eigenvalue as the trust value for each player to identify malicious players 
among all players.

\begin{figure}[htp]
  \centering
  \includegraphics[width=\linewidth]{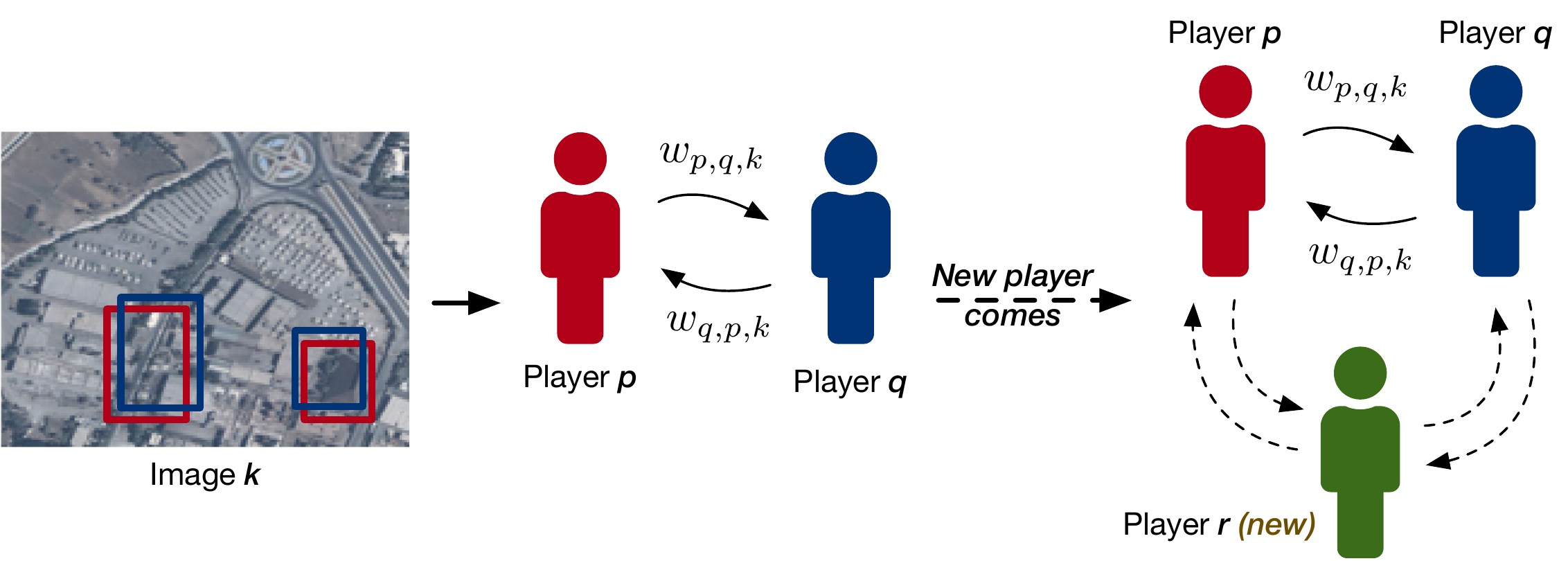}
  \caption{PRG for certain images: Assume player $p$ and $q$ are former players who have been
  evaluated as reliable players. Consider each of the player as a node in a graph,
  the $w_{p,q,k}$ is players' PRMR shown as a directed arc in the graph. 
  A new player is composed with former players in the graph as a game network.}
  \label{fig:graph}
\end{figure}

The model is established from image dependent perspective. For a certain image $k$,
considering a directed \emph{player rating graph (PRG)}
between players who tagged the image $k$. Each player is a node of PRG, 
as illustrated in Figure \ref{fig:graph}. 

\begin{definition}
\label{def:weightv}
Assuming the database stores $n$ different tags 
$g_1$, $g_2$, ..., $g_n$.
The system weight vector
$\mathbf{v} = (p(g_1)$, $p(g_2)$, ..., $p(g_n))^\top$ 
of all tags can be calculated by the following Equation \ref{eq:ptag}:
\begin{equation}
\label{eq:ptag}
p(g_i) = \frac{|g_i|}{\sum_{j=1}^{n}{|g_j|}}
\end{equation}
where $|g_i|$ is the count of $g_i$ in the system.
\end{definition}

\begin{definition}
\label{def:weightvk}
Assuming different tags 
$g_{r_1}$, $g_{r_2}$, ..., $g_{r_s}$ 
were tagged in
a certain image $k$, the image weight vector is a vector for image $k$ that is
composed by part of the system weight vector, which 
is denoted by 
$\mathbf{v}_k = $$(p(g_{r_1})$, $p(g_{r_2})$, $...$, $p(g_{r_s}))^\top$
with $r_i (i=1,2,...,s) \in \{1, 2, ..., n\}$, $r_i \neq r_j (i\neq j, j=1,2,...,s)$ and $s \leq n$.
\end{definition}

\begin{remark}\emph{
For instance, the system has 2 different images. 
The first image is tagged by two players. One is 
$g_1$, $g_2$, $g_5$ and another is $g_1$, $g_2$;
The second image is tagged by three players, their results are:
$g_1$, $g_2$, $g_5$;
$g_2$, $g_4$, $g_5$; 
$g_3$, $g_4$, $g_5$. 
Thus, the system currently has 5 different tags 
$g_1$, $g_2$, $g_3$, $g_4$, $g_5$.
Each tag has corresponding counts: $3, 4, 1, 2, 4$; 
Therefore the system weight vector is 
$(\frac{3}{14}$, $\frac{2}{7}$, $\frac{1}{14}$, $\frac{1}{7}$, $\frac{2}{7})^\top$;
the image weight vector of the first image is 
$(\frac{3}{14}$, $\frac{2}{7}$, $\frac{2}{7})^\top$
since the first image only is tagged by $g_1$, $g_2$, $g_5$, and
the image weight vector of the second image is the same as the system weight vector
since the second image is tagged by all exist tags.
}\end{remark}

\begin{lemma}
$p(g_i)$ holds the properties: 
\label{lemma:pg}
a) $0 \leq p(g_i)\leq 1$,
b) $\sum_{i=1}^{n}{p(g_{i})}=1$, and
c) $\sum_{i=1}^{s}{p(g_{r_i})} \leq 1$.
\end{lemma}

So far our player has two different type of inputs: \emph{the \hyperref[def:roi]{ROI}, 
and its tag vector}. To define the PRG edge weight, 
we introduce two input measurements in the subsequent Definition \ref{def:prmr} and \ref{def:pitc}.

\begin{definition}
\label{def:prmr}
The players ROI matching ratio (PRMR) is an importance measurement that measures the proportion of
two different \hyperref[def:roi]{ROI} intersection surface from player $p, q$ and 
the \hyperref[def:roi]{ROI} surface from player $p$ in a certain image $k$, 
which is denoted by the following formula:
\begin{equation}
\text{PRMR}(p, q, i, j, k) = \frac{| ROI_{p,i,k} \cap ROI_{q,j,k} | }
        {|ROI_{p,i,k}|}
\end{equation}
where $ROI_{p, i, k}$ is the $i$-th selected ROI from player $p$,
and $|ROI_{p, i, k}|$ is the surface area of $ROI_{p, i, k}$.
\end{definition}

\begin{lemma}
The following inequality holds:
\label{lemma:prmrrange}
\begin{equation}
\label{eq:prmrrange}
0 \leq \text{PRMR}(p, q, i, j, k) \leq 1
\end{equation}
\end{lemma}

\begin{definition}
\label{def:pitc}
The players input tag correlation (PITC) is an importance measurement that measures the proportion of
the covariance of two different \hyperref[def:tagv]{tag vectors $\mathbf{T}_{p,i,k}, \mathbf{T}_{q,j,k}$} from player $p, q$ 
and the covariance of $\mathbf{T}_{p,i,k}$ from player $p$ with itself under the
\hyperref[def:weightvk]{image weight vector $\mathbf{v}_k$}, which is denoted by the following formula:
\begin{equation}
\text{PITC}(p, q, i, j, k) = \frac{Cov(\mathbf{T}_{p,i,k}, \mathbf{T}_{q,j,k}; \mathbf{v}_k)}{Cov(\mathbf{T}_{p,i,k}, \mathbf{T}_{p,i,k}; \mathbf{v}_k)}
\end{equation}
where $Cov(\mathbf{X}, \mathbf{Y}; \mathbf{w})$ is the weighted covariance between $\mathbf{X}$ and $\mathbf{Y}$, which denoted by:
\begin{align}
\label{eq:cov}
& Cov(\mathbf{X}, \mathbf{Y}; \mathbf{w}) = \\
& \frac{\sum_{i=1}^{n}{w_i(x_i-\frac{1}{n}\sum_{i=1}^{n}{w_i x_i})(y_i-\frac{1}{n}\sum_{i=1}^{n}{w_i y_i})}}{\sum_{i=1}^{n}{w_i}}
\end{align}
with $\mathbf{X} = (x_1, x_2, ..., x_n)^\top, \mathbf{Y} = (y_1, y_2, ..., y_n)^\top, \mathbf{w} = (w_1, w_2, ..., w_n)^\top$.
\end{definition}

\begin{remark}\emph{
The definition of PRMR and PITC share the same intent for measuring asymmetric importance  
between player $p$ and player $q$ (namely how $p$ thinks of $q$).
}\end{remark}

\begin{remark}\emph{
The definition of PRMR is inspired by intersection over union (IoU),
a wide-used computer vision criteria also as known as Jaccard index \cite{real1996probabilistic, jaccard1901etude}.
statistically used to compare the similarity and diversity of sample sets. 
Differ from IoU, we only divided a single ROI surface area to guarantee the asymmetric property for directed graph weight.
}\end{remark}
\begin{remark}\emph{
The definition of PITC is inspired by the weighted pearson correlation coefficient \cite{pearson1895note},
which is a measure of the linear correlation between two variables. In our case, with the same intent of PRMR, 
we drop the
part of covariance of player $q$ in denominator to guarantee the asymmetric property for directed graph weight.
}\end{remark}
\begin{remark}\emph{
The PRMR and PITC both are not metrics of distance due to 
$\text{PRMR}(p, q, i, j, k)$ $\neq$ $\text{PRMR}(q, p, i, j, k)$ as well as
$\text{PITC}(p, q, i, j, k)$ $\neq$ $\text{PITC}(q, p, i, j, k)$.
}\end{remark}

\begin{lemma}
  The following inequality holds:
  \label{lemma:pitcrange}
  \begin{equation}
  \label{eq:pitcrange}
  -1 \leq \text{PITC}(p, q, i, j, k) \leq 1.
  \end{equation}
\end{lemma}

So far, we have enough techniques to define the edge weight of PRG.

\begin{definition}
  \label{def:edgeweight}
  For a certain image $k$, the edge weight of the PRG between player $p$ and $q$ is denoted 
  by the formula \ref{eq:weight}:
  \begin{equation}
  \label{eq:weight}
  w_{p,q,k} = 
  \sum_{j=1}^{n}{
  \sum_{i=1}^{m}{
    \text{PRMR}(p, q, i, j, k)
    \left(
      \text{PITC}(p, q, i, j, k) + 2
    \right)
  }}
  \end{equation}
  with player $p$ selected $m$ ROIs, player $q$ selected $n$ ROIs.
\end{definition}

The Perron-Frobenius theorem guarantees our goal can be drifted to the calculation of 
the adjacency matrix of PRG.
In consequence, one can use the normalized adjacency matrix by using formula \ref{eq:normalize}:

\begin{equation}
  \label{eq:normalize}
  \mathbf{A}_k = (a_{p,q,k}) = (\frac{w_{p,q,k}}{\sum_{q}{w_{p,q,k}}})
\end{equation}

where $k$ is the image indicator.

\begin{theorem}[Soundness]
  \label{theorem:property}
  The normalized adjacency matrix $A_k$ of PRG of a certain image $k$ is irreducible, real, 
  non-negative, and column-stochastic, with positive diagonal element.
\end{theorem}

\begin{remark}\emph{
From the proof (in Appendix \ref{appendix:proofs}) of property of positive diagonal elements, 
one can observe that the number ``2'' is 
a translation that guarantees $\text{PITC}(p, q, i, j, k)$ lies on closed interval $[1, 3]$ which helps us
prove this theorem successfully.
}\end{remark}

According to Perron-Frobenius theorem and 
Theorem \ref{theorem:property},
one can infer that there exists an uniqueness eigenvector 
$\mathbf{V}_k =$ $(\lambda_{1,k}$, $...$, $\lambda_{n,k})^\top$ of $\mathbf{A}_k$ (Perron vector),
with an uniqueness eigenvalue $\rho(\mathbf{A}_k)$ is the spectral radius of $\mathbf{A}_k$ (Perron root), such that:

\[
\mathbf{A}_k \cdot \mathbf{V}_k = \rho(\mathbf{A}_k) \cdot \mathbf{V}_k, \lambda_{i,k} > 0, \sum_{i=1}^{n}{\lambda_{i,k}} = 1.
\]

Therefore, we define the trust value of a player as following:

\begin{definition}
\label{def:tv}
A trust value $\lambda_{i,k}$ of player $i$ on image $k$ is a score that
equals to the $i$-th component of the Perron vector of the normalized PRG adjacency matrix $\mathbf{A}_k$.
\end{definition}

This definition represents the rating score from player $p$ to player $q$ for a certain image $k$, as same as the centrality
of the player $q$. With the trust value of players, we propose our classification algorithm:

\begin{algorithm}
  \SetAlgoLined
  \SetKwInOut{Input}{input}\SetKwInOut{Output}{output}
  \Input{New Player $p$,\\
         Trusted Player $p1, p2, ..., p_m$,\\
         Task Images $k_1, k_2, ..., k_{2n}$,\\
         Acceptance Threshold $\delta$}
  \Output{Reliability of Player $p$}
  \Begin{
    counter $\longleftarrow$ 0\\
    reliability $\longleftarrow$ false\\
    \For{$k \in [k_1, k_2, ..., k_{2n}]$}{
      \If{$k$ is tagged image}{
        calculate $\lambda_{p,k}, \lambda_{p_1,k}, ..., \lambda_{p_m,k}$\\
        \If{$\lambda_{p,k} \geq \frac{1}{m}\sum_{i=1}^{m}{\lambda_{p_i, k}}$}{
          counter $\longleftarrow$ counter $+ 1$
        }
      }
    }
    \If{counter $\geq \delta$}{
      reliability $\longleftarrow$ true
    }
  }
  \caption{Malicious Player Detection}
  \label{algo:malicious}
\end{algorithm}

\begin{remark}\emph{
The criterion of classifying new players performs the action that 
the trust value of a new player should not be less than the mean value of overall trust value of players on image $k$, 
which means the tagging performance of new player should not be worse than result performance of former players.
The acceptance threshold is a customizable parameter that can be set beforehand.
For instance, if $\delta = 1$, the new player only needs to pass one singular image of all tagged images; 
if $\delta = n$ (half images of the task), the new player has to pass all tagged images, which makes
the system unbreakable if the system is initialized by a trusted group.
}\end{remark}

Note that sometimes new player carries new tags into the system. It will influence the 
\hyperref[def:tagv]{tag vector} calculation and
cause the weight not computable due to 
the inequal dimensions of the \hyperref[def:tagv]{tag vector} of new player and old player.
A solution for this issue is proposed in the following steps:

\begin{itemize}
  \item If a new player does not provide new tag: Directly perform the calculation with the algorithm;
  \item If a new player carries new tags only: Directly drop them because they are unreliable;
  \item If a player carries both selected and new tags: 
    a) Perform the calculation with the algorithm without new tags;
    b) Merge and update all weight vector $v$ via formula \ref{eq:weight} if the player is reliable;
    c) Otherwise drop and mark the result as unreliable.
\end{itemize}

\subsection{Disaster Evaluation Model}

The idea of stochastic pooling \cite{ciresan2011flexible, krizhevsky2012imagenet, ciregan2012multi} 
is applied to define our Disaster Evaluation Model.
For a monitoring region at time $t$, we address the \emph{DEM} 
through disaster level definition as follows:

\begin{definition}
  \label{def:dl}
  A monitor region is composed by images $k_1, ..., k_n$.
  Each image exists $r_{k_i}$ number of ROIs with $i=1, ..., n$,
  and each ROI is tagged with tags $g_1, ..., g_m$.
  The \emph{disaster level} $\Delta$ of a monitor region is calculated by the following:
  \begin{equation}
  \label{eq:dl}
  \Delta = 
    \sum_{j=1}^{m} \left(
    p(g_j)
    \frac{
      \sum_{g_j} |ROI|
    }{
      \sum_{i = 1}^{n} |k_i|
    }
    \right)
  \end{equation}
  where $|ROI|$ is the surface area of a ROI, $\sum_{g_j} |ROI|$ means accumulated surface area
  of all ROIs that tagged by $g_j$, and $|k_i|$ is the surface area of image $k_i$.
\end{definition}

\begin{remark}\emph{
The disaster level is defined as a weighted area coverage. 
The $\frac{\sum_{g_j} |ROI|}{\sum_{i = 1}^{n} |k_i|}$ is a surface ratio of the ROI over monitoring area,
and the $p(g_i)$ is the correcponding weight of the ratio.
}\end{remark}

\begin{theorem}[Denseness]
\label{theorem:complete}
The disaster level $\Delta$ is dense in internal $[0, 1]$.
\end{theorem}

With Theorem \ref{theorem:complete}, one can calculate the disaster level for 
a monitoring region according to Equation \ref{eq:dl}.
An area gains higher disaster if the disaster level is closer to 1, and vise versa.

\subsection{Model Initialization}
\label{sec:modelinit}

Due to the lack of users in the very beginning, 
cold start is a common problem in such human computation system. 
This issue is nornally solved
by hiring people to create data manually.
In this system, we only have to consider one system initialization issue of cold start.

\begin{figure}[!htbp]
  \centering
  \includegraphics[width=\linewidth]{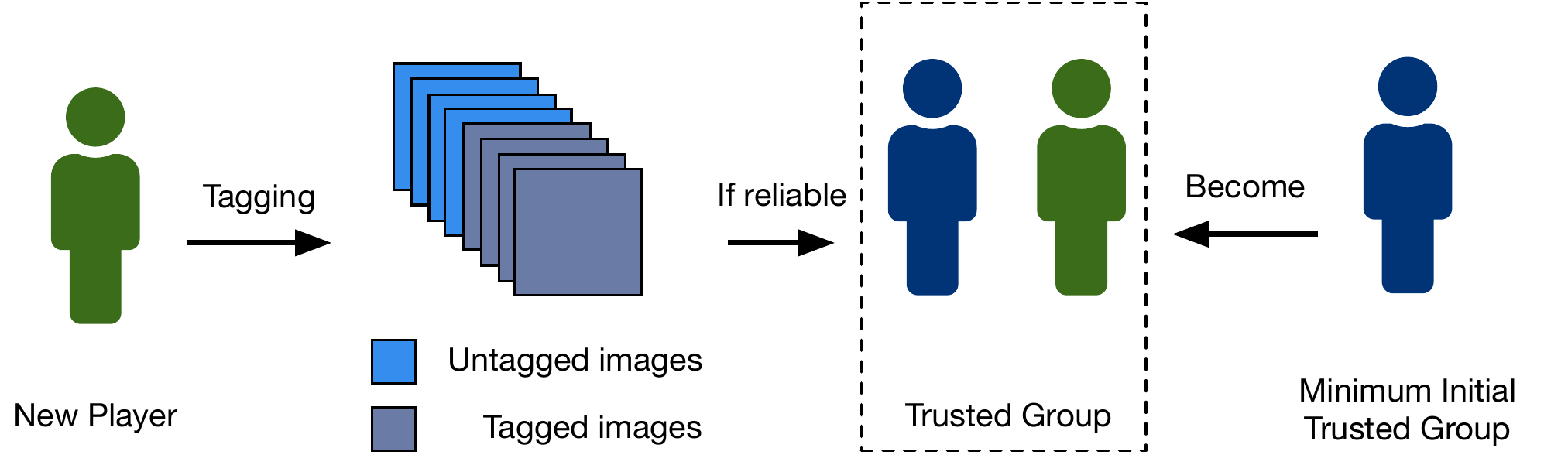}
  \caption{Initialization of PRM}
  \label{fig:cold}
\end{figure}

The issue appears in the PTG. 
To initialize the whole system, we need to address an \emph{initial trusted group} for PTG
who shall tag enough initial trusted results as well as a fixed predefined tag list 
(containing all of the most important keywords that need to be monitored)
for PTG and then assign the tagged images to new upcoming players. 
Once a new player is included in the trusted group, all the relevant result from this player 
will be considered as reliable. 
The trusted group and available dataset grows with gradually growing number 
of reliable players and their reliable tags, 
as shown in Figure \ref{fig:cold}.

Thus, we have only one issue regarding the minimum number of the initial trusted group.
Our PRM is based on graph centrality calculation, which means we need a (at least) two dimensional matrix
to perform the overall model calculation. Hence, with the new player, \emph{the minimum number of the initial trusted group is 1}.
Then the initial trusted group (one person) with the new player form a two dimensional adjacency matrix that makes the model
computable. For larger initial trusted groups, the \hyperref[def:tv]{trust value} can be simply initialized 
to $\frac{1}{n}$ with $n$ is the number of initial trusted group.

\section{Discussion}

We have described the system architecture and 3 core models for task generation, 
malicious player detection and disaster level evaluation.
Malicious player detection is essentially 
a classification problem in which 
our system determines the reliability of a new player
based on the \hyperref[def:tv]{trust value}. 
In this section, 
we would like to discuss some issues for the future work.

\subsection{Simulated evaluation}

To evaluate our model, a typical classification model 
performance evaluation metric is 
receiver operating characteristic (ROC) curve \cite{hanley1982meaning}, 
which 
plots True Positive (on the y-axis) against False Positive and 
the ideal \emph{surface under the ROC curve} is 1.
Nevertheless, before we test the system with real users, 
one can generate a reasonable random dataset 
to test the performance of 
our classification model (PRM).

Our player has two different types of inputs: 
\emph{the ROI and its tag vector}.
For a reasonable player data entity, 
one has to define the ROI selection and its corresponding tag vector.
To generate reasonable ROI for simulating real user behavior, 
we would like to discuss a desktop target click behavior first.

\begin{figure}
  \centering
  \includegraphics[width=\linewidth]{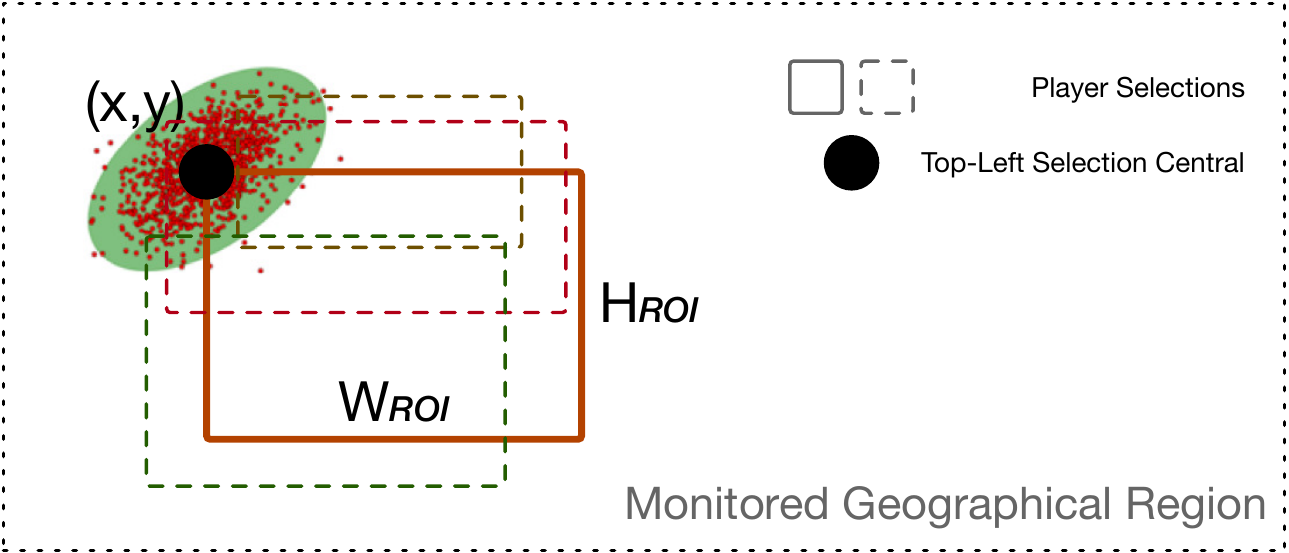}
  \caption{An example of ROI simulation which can be used in the system evaluation.}
  \label{fig:evaluation}
\end{figure}

The target click behavior on a screen has been explored for years \cite{mackenzie1992fitts, bi2013ffitts}.
It has been modeled and proved that 
the distribution of click behavior for a certain point 
satisfy Gaussian distribution \cite{goodman1963statistical}.
Thus, from frequency statistic view, the actual ROI(s) certainly exists.
No matter where the user starts,
according to the Fitts Law and FFitts Law, 
the starting click point should follows normal distribution around the actual point,
as shown in Figure \ref{fig:evaluation}. 
Similarly, the end point of the selection of ROI(s) should also follows a normal distribution.

Therefore, to generate ROI(s), let $(x, y)$ as the player ROI starting point, $(H_{ROI}$, $W_{ROI})$ as the height
and width pair of this ROI, then we generate noise for the ROI starting points and landing points: 
$(x+\epsilon, y+\epsilon), (H_{ROI}+\epsilon, W_{ROI}+\epsilon)$ where $\epsilon \sim N(0, \delta)$.
For the parameter $\delta$, one can use maximum likelihood estimator \cite{johansen1990maximum}
to perform the inference for all manually ROI selection samples from initial trusted group.

The generation of tag vector for a certain image is simpler than ROI's.
A randomly pick from initial trusted group is sufficient for the simulation case 
because these tags are trusted results and a partially randomly selection already introduced the noise in this case.

Eventually, one can apply this random dataset 
to evaluate \emph{surface under the ROC curve} 
as an indication of the overall performance
(the model may show good performance if the surface approximate to 1).

\subsection{Data leakage and information loss}
\label{sec:ethical}

\begin{figure}
  \centering
  \includegraphics[width=\linewidth]{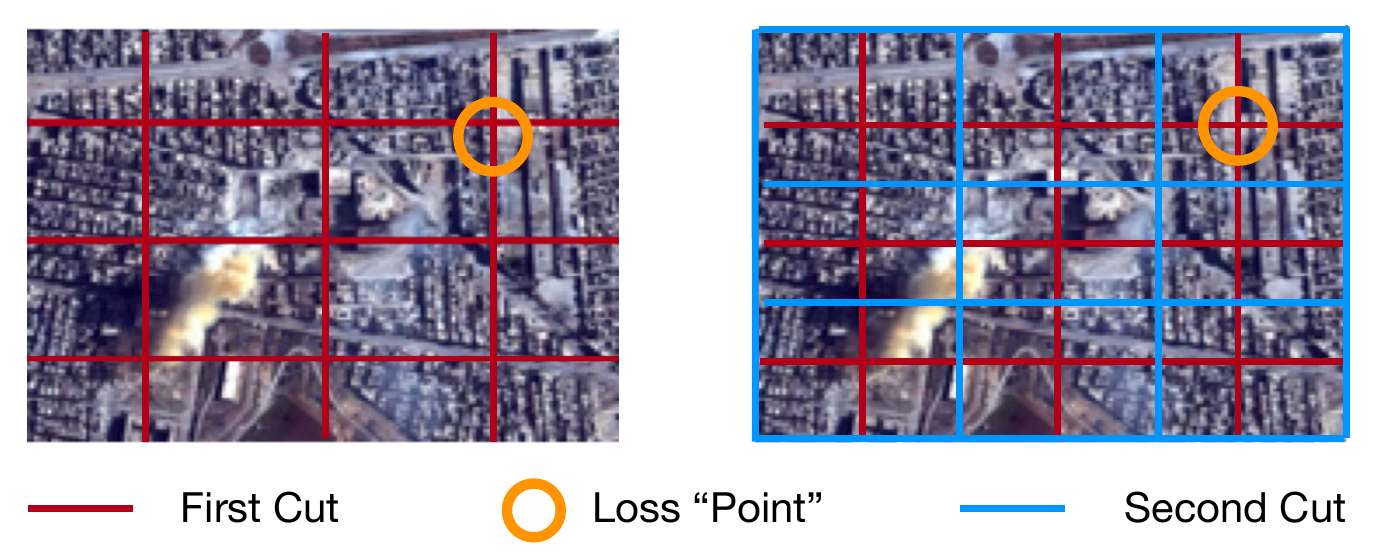}
  \caption{
      Information loss may occur on the intersection lines; 
      a possible solution is to perform a ``half shifting'' cut.
  }
  \label{fig:information_loss}
\end{figure}

In order to prevent leakage of data to malicious players, 
we intentionally cut original satellite images into small segmentations.
However, this method may cause information loss if some important ROIs are 
located at the intersection of two dividing lines.
A possible solution is to consider ``half shifting'' cut,
as shown in Figure \ref{fig:information_loss}.

\subsection{Limitations}

\paragraph{Outdated Evaluation}

Our PRG network is based 
on image dependent perspective, that leads,
each calculated \hyperref[def:dl]{disaster level} may become invalid if the region image is outdated.
We assume the satellite takes pictures for the monitoring area between intervals. 
However, our model only calculates the disaster level at a unique moment, 
which means the disaster level needs transvaluation when a new image is generated.
If none of the new images gets evaluated, then the disaster level will not be updated.
The disaster level of a certain region over time is essentially a non-stationary process \cite{brockwell2013time} 
time series prediction method \cite{kuznetsov2015learning} can be applied
on the disaster level time series.

\paragraph{Game Playability}

Considering the fact that
most parts of the earth are lake, forest, 
desert and so on, during the game playing, players
may meet the situation that there is no available 
ROI in several continuous rounds. Obviously, it will decrease the playability and enjoyment of the game.
A possible solution is pre-filtering these images from the image database.

\section{Conclusion}

In this paper, we explored a GWAPs-based disaster monitoring system.
We firstly proposed a \emph{player rating model} based on 
eingenvalue centralities to calculate the trust value of a player.
And then we proposed an algorithm for malicious user detection. 
As justification, we proved the mathematical correctness of this model.
We then calculate the regional disaster level
in the \emph{disaster evaluation model}.
We also deal with the general problem of system cold start by introducing 
the method of image half shifting cut.
Our system design can also applied to other similar human computation systems.
Furthermore, we discussed theoretical evaluation criteria for this system,
and then addressed corresponding solutions for the issues of 
data leakage, information loss and game playability.

\section*{Acknowledgment}

The authors would like to thank Prof. Fran\c{c}ois Bry and Prof. Andreas Butz for
their valuable input;
we also thank colleague Yingding Wang for his inspiration on 
system design, algorithm rationalizations as well as system evaluations.
Finally, we also thank Huimin An for his inspiration on Bayesian perspective that
helps us handling human inputs with new tags successfully.

\nocite{*}
\bibliographystyle{IEEEtran}
\bibliography{IEEEabrv,ou2019gwap}

\appendix

\subsection{Examples of database scheme}
\label{appendix:scheme}

With Definition \ref{def:roi}, the players of our system are able to 
select ROIs for each image as well as capable of select tags for each ROI.
Thus, the \emph{tasks} field in PlayerDB is an array object, 
stores each player image result with an assigned identifier.

\begin{listing}[H]
\begin{lstlisting}
[{
  "player_id": "E3A6F124-4A6C-4C6E-B7F1-F8BC9A7381CC",
  "tasks": [{
      "image_id": "3A21E99E-F074-454B-A590-8D8C5ABD8E77",
      "image_at": "2017-07-31 11:28:40",
      "reliable": true,
      "ROIs": [{ "x": 103, "y": 121, "height": 56, "width": 78,
          "tags": ["burning building", "explosion"]
      }]
  }]
}, ...]
\end{lstlisting}
\caption{An example of PlayerDB data scheme}
\label{lst:playerdb}
\end{listing}

Each object in the \emph{tasks} array has a field \emph{reliable}, 
which indicates the reliability for this object task;
Each object also contains a \emph{ROIs} field, which is an array object that 
contains the player inputs for this object image;
Each ROI object in the ROIs field has four properties that describes 
the ROI geometric location: \emph{x, y, height, width}, and 
also a \emph{tags} array field that describes the input tags for this image 
from this player.
For \emph{tags} field, game players can select the related tags for each ROI, 
and stores in this array. 

\begin{listing}
\begin{lstlisting}
[{
  "region_id": "FBEB6204-0B94-4811-94F0-9DDC5FBBE6D8",
  "history": [{
    "image_id": "3A21E99E-F074-454B-A590-8D8C5ABD8E77",
    "image_at": "2017-07-31 11:28:40",
    "ROIs": [{
      "x": 103, "y": 121, "height": 56, "width": 78,
      "tags": ["burning building", "explosion"]
    }]
  }]
}, ...]
\end{lstlisting}
\caption{An example of ResultDB data scheme}
\label{lst:resultdb}
\end{listing}

\subsection{Proof of Lemma \ref{lemma:pg}}

\begin{IEEEproof}
\emph{a)} According to the Definition \ref{def:weightv}, 
$|g_i|$ is non-negative, then $0 \leq |g_i| \leq \sum_{j=1}^{n}{|g_i|}$.
Thus, we have $0 \leq p(g_i) \leq 1$.
\emph{b)} $\sum_{i=1}^{n} p(g_i) = \sum_{i=1}^{n} \frac{|g_i|}{\sum_{j=1}^{n}{|g_j|}} = 1$.
\emph{c)} $\sum_{i=1}^{s} p(g_{r_i}) \leq \sum_{i=1}^{n} p(g_{r_i}) \leq 1$.
\end{IEEEproof}

\subsection{Proof of Lemma \ref{lemma:prmrrange}}

\begin{IEEEproof}
  According to the Definition of \hyperref[def:roi]{ROI}, $| ROI_{p,i,k} \cap ROI_{q,j,k} |$ 
  can archive its maximum value only and only if $ROI_{p,i,k} =  ROI_{q,j,k}$ 
  as well as its minimum value only and only if $ROI_{p,i,k}$ has no intersection with $ROI_{q,j,k}$.
  Thus:

\begin{equation}
\begin{split}
0 =& \frac{0}{|ROI_{p,i,k}|} \leq \text{PRMR}(p, q, i, j, k)  \\
\leq& \frac{| ROI_{p,i,k} \cap ROI_{p,i,k} | }{|ROI_{p,i,k}|} = \frac{|ROI_{p,i,k}|}{|ROI_{p,i,k}|} = 1.
\end{split}
\end{equation}
\end{IEEEproof}

\subsection{Proof of Lemma \ref{lemma:pitcrange}}

\begin{IEEEproof}
We know that the weighted Pearson Correlation Coefficient \cite{pearson1895note} lies on $[-1, 1]$, i.e.
\[
  -1 \leq \frac{Cov(\mathbf{T}_{p,i,k}, \mathbf{T}_{q,j,k}; \mathbf{v}_k)}{\sqrt{Cov(\mathbf{T}_{p,i,k}, \mathbf{T}_{p,i,k}; v_k)Cov(\mathbf{T}_{q,j,k}, \mathbf{T}_{q,j,k}; \mathbf{v}_k)}} \leq 1
\]
To prove Equation \ref{eq:pitcrange}, we have to show:
\begin{multline}
  \frac{Cov(\mathbf{T}_{p,i,k}, \mathbf{T}_{q,j,k}; \mathbf{v}_k)}{Cov(\mathbf{T}_{p,i,k}, \mathbf{T}_{p,i,k}; \mathbf{v}_k)} \leq |Cov(\mathbf{T}_{q,j,k}, \mathbf{T}_{q,j,k}; \mathbf{v}_k) \\ 
  \sqrt{Cov(\mathbf{T}_{p,i,k}, \mathbf{T}_{p,i,k}; \mathbf{v}_k)Cov(\mathbf{T}_{q,j,k}, \mathbf{T}_{q,j,k}; \mathbf{v}_k)}| \leq 1
\end{multline}
and
\begin{multline}
  \frac{Cov(\mathbf{T}_{p,i,k}, \mathbf{T}_{q,j,k}; v_k)}{Cov(\mathbf{T}_{p,i,k}, \mathbf{T}_{p,i,k}; \mathbf{v}_k)} \geq -|Cov(\mathbf{T}_{q,j,k}, \mathbf{T}_{q,j,k}; \mathbf{v}_k) \\
  \sqrt{Cov(\mathbf{T}_{p,i,k}, \mathbf{T}_{p,i,k}; v_k)Cov(\mathbf{T}_{q,j,k}, \mathbf{T}_{q,j,k}; \mathbf{v}_k)}| \geq -1
\end{multline}
Then we need to show:
\begin{equation}
0 \leq Cov(\mathbf{T}_{p,i,k}, \mathbf{T}_{p,i,k}; \mathbf{v}_k)Cov(\mathbf{T}_{q,j,k}, \mathbf{T}_{q,j,k}; \mathbf{v}_k)^3 \leq 1
\end{equation}
Considering $\mathbf{T}_{p,i,k}, \mathbf{T}_{q,i,k}$ are described in general, with Equation \ref{eq:cov}, 
we only need to show (\emph{$s$ is an vector components index} instead of exponential):
\begin{equation}
\label{eq:statement}
\begin{split}
0 &\leq Cov(\mathbf{T}_{p,i,k}, \mathbf{T}_{p,i,k}; \mathbf{v}_k) \\
  &= 
\frac{
  \sum_{s=1}^{n}{
    \mathbf{v}_{k}^s
    \left(\mathbf{T}_{p,i,k}^s - \frac{1}{n}\sum_{s=1}^{n}{\mathbf{v}_{k}^s \mathbf{T}_{p,i,k}^s}\right)^2
  }
}{
  \sum_{s=1}^{n}{\mathbf{v}_{k}^s}
} \leq 1
\end{split}
\end{equation}
According to the definition of \hyperref[def:tagv]{tag vector} and \hyperref[def:weightvk]{image weight vector},
the components of $\mathbf{T}_{p,i,k}$ are either 1 or 0, 
the components of $\mathbf{v}_k$ lies on $[0, 1]$, with Lemma \ref{lemma:pg}, we have:
\begin{equation}
0 \leq \left(\mathbf{T}_{p,i,k}^s - \frac{1}{n}\sum_{s=1}^{n}{\mathbf{v}_{k}^s \mathbf{T}_{p,i,k}^s}\right)^2 \leq 1
\end{equation}
Therefore,
\begin{equation}
0 = \frac{
  \sum_{s=1}^{n}{
    v_{k}^s \cdot 0
  }
}{
  \sum_{s=1}^{n}{v_{k}^s}
} 
\leq Cov(T_{p,i,k}, T_{p,i,k}; v_k) \leq
\frac{
  \sum_{s=1}^{n}{
    v_{k}^s \cdot 1
  }
}{
  \sum_{s=1}^{n}{v_{k}^s}
} = 1
\end{equation}
which proves Equation \ref{eq:statement}.
\end{IEEEproof}

\subsection{Proof of Theorem \ref{theorem:property}}
\label{appendix:proofs}
\begin{IEEEproof}

\emph{Irreducibility} As shown in Figure \ref{fig:graph}, for a certain image $k$, 
the PRG is strong connected because the
player who selected ROIs in image $k$ has a direct connection to any other player who also selected ROIs in image $k$ 
(the edge weight is well defined according to Equation \ref{eq:weight}).
Thus, since $A_k$ is an normalized strong connected PRG adjacency matrix, which proves $A_k$ is irreducible.
  
\emph{Real elements} With Lemma \ref{lemma:prmrrange} and \ref{lemma:pitcrange}, each part of the Equation \ref{eq:weight} 
are real number. Thus, of course, the matrix $A_k$ elements are calculated by Equation \ref{eq:normalize} that are real elements.
  
\emph{Non-negative elements} With Lemma \ref{lemma:prmrrange} and \ref{lemma:pitcrange}, we have:

\begin{align}
& w_{p,q,k} \\
&= 
\sum_{j=1}^{n}{
\sum_{i=1}^{m}{ \left(
  \text{PRMR}(p, q, i, j, k)
  \left(
    \text{PITC}(p, q, i, j, k) + 2
  \right)
\right)}} \\
&\geq
\sum_{j=1}^{n}{
\sum_{i=1}^{m}{ \left(
  0 \cdot
  \left(
    -1 + 2
  \right)
\right)}} = 0
\end{align}

Thus, $w_{p,q,k}$ has its lower bound when 
$\text{PRMR}(p, q, i, j, k) = 0$ (for $all i=1,...,m; j=1,...,n$) 
and $\text{PITC}(p, q, i, j, k) = -1 $(for all $i=1,...,m; j=1,...,n)$. 
Meanwhile,

\begin{align}
& w_{p,q,k} \\
&= 
\sum_{j=1}^{n}{
\sum_{i=1}^{m}{ \left(
  \text{PRMR}(p, q, i, j, k)
  \left(
    \text{PITC}(p, q, i, j, k) + 2
  \right)
\right)}} \\
&\leq 
\sum_{j=1}^{n}{
\sum_{i=1}^{m}{ \left(
  1 \cdot
  \left(
    1 + 2
  \right)
\right)}} = 3mn
\end{align}
  
and $w_{p,q,k}$ has its upper bound when $\text{PRMR}(p, q, i, j, k) = 1 (\text{for all} i=1,...,m; j=1,...,n)$ 
and $\text{PITC}(p, q, i, j, k) = 1 (\text{for all} i=1,...,m; j=1,...,n)$.
    
\emph{Positive diagonal elements} According to Lemma \ref{lemma:pitcrange}, 
the diagonal elements can be formalized by follows:
  
\begingroup
\allowdisplaybreaks
\begin{align}
&w_{p,p,k} \\
&= 
\sum_{j=1}^{m}{
\sum_{i=1}^{m}{ 
  \left(
    \text{PRMR}(p, p, i, j, k)
    \left(
      \text{PITC}(p, p, i, j, k) + 2
    \right)
  \right)
}} \\
&\geq \sum_{j=1}^{m}{
\sum_{i=1}^{m}{ \left(
  \frac{| ROI_{p,i,k} \cap ROI_{p,j,k} | }{|ROI_{p,i,k}|}
  \left(
    -1 + 2
  \right)
\right)}} \\
&= \sum_{j=1}^{m}{
\sum_{i=1}^{m}{
  \frac{| ROI_{p,i,k} \cap ROI_{p,j,k} | }{|ROI_{p,i,k}|}
}}\\
&= \sum_{i=j}{\frac{| ROI_{p,i,k} \cap ROI_{p,j,k} | }{|ROI_{p,i,k}|}} \\
&+ \sum_{i\neq j}{\frac{| ROI_{p,i,k} \cap ROI_{p,j,k} | }{|ROI_{p,i,k}|}}\\
&\geq \sum_{i=j}{\frac{| ROI_{p,i,k} \cap ROI_{p,j,k} | }{|ROI_{p,i,k}|}}\\
&= \sum_{i=1}^{m}{\frac{| ROI_{p,i,k} \cap ROI_{p,i,k} | }{|ROI_{p,i,k}|}} \\
&= \sum_{i=1}^{m}{\frac{| ROI_{p,i,k}|}{|ROI_{p,i,k}|}} = m > 0
\end{align}
\endgroup
  
\emph{Column stochastic} according to the definition of matrix $A$, the sum of the column
elements are:

\begin{equation}
\begin{split}
\sum_{q}{a_{p,q,k}} = \sum_{q}{ \frac{w_{p,q,k}}{ \sum_{q}{w_{p,q,k}} }} = \frac{\sum_{q}{w_{p,q,k}}}{\sum_{q}{w_{p,q,k}}} = 1
\end{split}
\end{equation}

\end{IEEEproof}

\subsection{Proof of Theorem \ref{theorem:complete}}
\label{appendix:proofs}
\begin{IEEEproof}

According to the Definition \ref{def:dl} and Lemma \ref{lemma:pg}, it is trivial to show $\sup\Delta = 1$ and $\inf\Delta = 0$,
since $p(g_i)$ lies in $[0, 1]$ and $\sum_{g_j} |ROI| \leq \sum_{i = 1}^{n} |k_i|$.

The rest of the proof will prove $\forall \Delta_p < \Delta_q$,
there exist $\Delta_r$ such that $\Delta_p < \Delta_r < \Delta_q$.

We may assume two monitored region $\Delta_p$ has $m_q$ tags and $\Delta_q$ has $m_q$ tags where $m_p + 2 = m_q$,
which indicates that there exists two tags $g_\alpha$ and $g_\beta$ are not appeared in $\Delta_p$ 
but in $\Delta_q$.
Let $m_r = m_p + 1$, i.e. one of $g_\alpha$ and $g_\beta$ appear in region $\Delta_r$,
thus we have:

\begin{align}
\Delta_p 
&= \sum_{j=1}^{m_p} \left(
  p(g_j)
  \frac{
    \sum_{g_j} |ROI|
  }{
    \sum_{i = 1}^{n} |k_i|
  }
  \right) \\
&< \sum_{j=1}^{m_p+1} \left(
  p(g_j)
  \frac{
    \sum_{g_j} |ROI|
  }{
    \sum_{i = 1}^{n} |k_i|
  }
  \right)
= \Delta_r \\
&< \sum_{j=1}^{m_p+2} \left(
p(g_j)
\frac{
  \sum_{g_j} |ROI|
}{
  \sum_{i = 1}^{n} |k_i|
}
\right)
= \Delta_q
\end{align}

\end{IEEEproof}

\end{document}